\documentclass[modern]{aastex63}

\hypersetup{linkcolor=red, citecolor=cyan,filecolor=cyan,urlcolor=blue}

\usepackage{dcolumn}

\shorttitle{Evidence for a nonlinear damping model for coronal loop oscillations}
\shortauthors{I. Arregui}

\begin{document}

\title{Bayesian evidence for a nonlinear damping model for coronal loop oscillations}
\email{iarregui@iac.es}
\author[0000-0002-7008-7661]{I\~nigo Arregui}
\affiliation{Instituto de Astrof\'{\i}sica de Canarias, V\'{\i}a L\'actea S/N, E-38205 La Laguna, Tenerife, Spain}
\affiliation{Departamento de Astrof\'{\i}sica, Universidad de La Laguna,
E-38206 La Laguna, Tenerife, Spain}

\begin{abstract}
Recent observational and theoretical studies indicate that the damping of solar coronal loop oscillations depends on the oscillation amplitude. We consider two mechanisms, linear resonant absorption and a nonlinear damping model. We confront theoretical predictions from these models with observed data in the plane of observables defined by the damping ratio and the oscillation amplitude. The structure of the Bayesian evidence in this plane displays a clear separation between the regions where each model is more plausible relative to the other. There is qualitative agreement between the regions of high marginal likelihood and Bayes factor for the nonlinear damping model and the arrangement of observed data. A quantitative application to 101 loop oscillation cases observed with SDO/AIA results in the marginal likelihood for the nonlinear model being larger in the majority of them. The cases with conclusive evidence for the nonlinear damping model outnumber considerably those in favor of linear resonant absorption. 
\end{abstract}



\section{Motivation} 

Damped transverse oscillations of solar coronal loops have been the subject of intense observational and theoretical research in the past two decades since their discovery using the Transition Region and Coronal Explorer (TRACE)  by \cite{aschwanden99} and \cite{nakariakov99}. Further observations with instrumentation onboard the Solar Dynamics Observatory (SDO) and the Solar TErrestrial RElations Observatory (STEREO) produced an increase in the amount and quality of the data. This has led to the creation of catalogs containing oscillation properties for a large number of events \citep{goddard16a,nechaeva19}. The analysis of these events shows an empirical relationship between the measured damping ratio and the oscillation amplitude \citep{goddard16b}.   

The observed damping has been widely attributed to the mechanism of resonant absorption \citep{ruderman02,goossens02a}. The numerical experiments by \cite{magyar16} show that for small oscillation amplitudes linear resonant absorption provides a good approximation to the damping properties.  For larger amplitudes the nonlinear evolution results in a significantly faster damping than predicted by linear theory. Recent analytical developments by \cite{vd2021}, using a mathematical description in terms of Els\"{a}sser variables,  show that the nonlinear damping time of impulsively excited standing kink modes in coronal loops is inversely proportional to the oscillation amplitude. 

Assessing the damping mechanism(s) operating in coronal loop oscillations is crucial to advance in coronal seismology and wave heating investigations. In this study, we confront the theoretical predictions from linear resonant damping and from the nonlinear damping model by \cite{vd2021} with observed loop oscillation properties in the catalog compiled by \cite{nechaeva19}. The aim is to quantify the evidence for the nonlinear damping model relative to the evidence for resonant absorption in explaining the damping of coronal loop oscillations.

\section{Damping Models} 

The damping of coronal loop oscillations has been widely interpreted in terms of resonant absorption (RA) of standing kink waves in radially inhomogeneous flux tube models \citep[see e.g.,][]{goossens06,goossens11}. The mechanism is based on the transfer of wave energy from large to small spatial scales in the radial direction. Under the thin tube and thin boundary approximations, the ratio of the damping time $\tau_{\rm d}$ to the oscillation period $P$ is given by the analytical expression \citep{ruderman02,goossens02a}

\begin{equation}\label{eq:ra}
\frac{\tau_{\rm d}}{P}\|_ {M_{\rm RA}} = \mathcal{F}\, \frac{\zeta+1}{\zeta-1}\, \frac{R}{l},
\end{equation}
with $\zeta=\rho_{\rm i}/\rho_{\rm e}$ the ratio of internal to external density, $ l/R$ the length of the non-uniform layer at the boundary of the waveguide with radius $R$, and $\mathcal{F}=2/\pi$ for a sinusoidal variation of density over the non-uniform layer. The predictions from the damping model $M_{\rm RA}$ given by Eq.~(\ref{eq:ra}) for the observable damping ratio are determined by the parameter vector $\mbox{\boldmath$\theta$}_{\rm RA}=\{\zeta, l/R\}$. 

A recent analytical investigation by \cite{vd2021} has shown that the nonlinear (NL) damping time of standing kink waves is inversely proportional to the oscillation amplitude. The nonlinear evolution of the dynamics produces an energy transfer to small scales in the radial and azimuthal directions. Using a formalism based on the use of Els\"{a}sser variables, \cite{vd2021} derive an analytical expression for the damping ratio in the inertial regime of the turbulent cascade given by 

\begin{equation}\label{eq:nl}
\frac{\tau_{\rm d}}{P}\|_{M_{\rm NL}} = 20\sqrt{\pi}\,\frac{1}{2\pi a}\, \frac{1+\zeta}{\sqrt{\zeta^2 -2\zeta+97}},
\end{equation}
with $ a=\eta/R$ the ratio of the displacement $\eta$ to the loop radius. The predictions from the damping model $M_{\rm NL}$ given by Eq.~(\ref{eq:nl}) for the observable damping ratio, for known oscillation amplitude, are determined by the parameter vector $\mbox{\boldmath$\theta$}_{\rm NL}=\{R, \zeta\}$.  

\section{Analysis and results} 

Our judgement on the evidence in favor of any of the two models with respect to the other in explaining a set of observations is based on the use of Bayesian reasoning \citep[see e.g.,][]{jaynes03,gregory05,lindley14}. The philosophy and methodology of this approach and early applications to solar coronal seismology are discussed in \cite{arregui18}.

As first noticed by \cite{goddard16b}, when the damping ratio for a large number of loop oscillation events is plotted against their oscillation amplitude, the data are scattered forming a cloud with a triangular shape. Larger amplitudes correspond in general to smaller damping ratio values and vice versa (see Figure 2 in \citealt{goddard16b} or Figure 6 (bottom right) in \citealt{nechaeva19}).  In our analysis, we quantify the relative ability of the two considered damping models to explain the distribution of the data in the plane of observables defined by the damping ratio and the oscillation amplitude. 

In the first part of the analysis, we calculate the marginal probability of the data for each model and compute their ratio, the Bayes factor, over a two-dimensional synthetic data space. This gives a bird's eye view of the general structure of the evidence.  In the second part of the analysis, Bayes factors are computed for the 101 loop oscillation events in the catalog by \cite{nechaeva19} with information about the oscillation amplitude.

\subsection{Structure of the evidence}\label{sec:structure}

Given a model $M$ with parameter vector $\mbox{\boldmath$\theta$}$ proposed to explain observed data $D$,  a relational measure of the quality of the model derives from the integral of the joint distribution $p(\mbox{\boldmath$\theta$},D|M)$ over the full parameter space 

\begin{equation}\label{eq:ml}
p(D|M) = \int_{\mbox{\scriptsize\boldmath$\theta$}} p(\mbox{\boldmath$\theta$}, D| M) \, d\mbox{\boldmath$\theta$} = \int_{\mbox{\scriptsize\boldmath$\theta$}} p(D|\mbox{\boldmath$\theta$},M) \, p(\mbox{\boldmath$\theta$}|M) \, d\mbox{\boldmath$\theta$}.
\end{equation} 
This is the so-called marginal likelihood. In this expression, $p(\mbox{\boldmath$\theta$}|M)$ is a prior distribution over the parameter space and $p(D|\mbox{\boldmath$\theta$},M)$ is the likelihood of obtaining a specific data realisation as a function of the parameter vector.  This measure of evidence is relational because a relation is examined to quantify how well the data $D$ are predicted by the model $M$.

To apply Eq.~(\ref{eq:ml}) to the damping models $M_{\rm NL}$ and $M_{\rm RA}$ a two-dimensional grid over synthetic data space $\mathcal{D}=(\eta, \tau_{\rm d}/P)$ is constructed. The grid covers the ranges in oscillation amplitude and damping ratio in the observational and theoretical studies by \cite{nechaeva19} and \cite{vd2021}. Possible data realisations over the grid are generated using the theoretical predictions given by Eqs.~(\ref{eq:ra}) and (\ref{eq:nl}) for models $M_{\rm RA}$ and $M_{\rm NL}$, respectively. 
Under the assumption of a Gaussian likelihood function and adopting an error model for the damping ratio alone

\begin{equation}\label{eq:like}
p(\mathcal{D}|\mbox{\boldmath$\theta$}, M_{\rm RA,NL}) =\frac{1} {\sqrt{2\pi} \sigma} 
\exp \Bigg\{-\frac{\left[\frac{\tau_{\rm d}}{P} - \frac{\tau_{\rm d}}{P}(\mbox{\boldmath$\theta_{}$})\|_{M_{\rm RA,NL}}\right]^2}{2\sigma^2}\Bigg\},
\end{equation}
with $\sigma$ the uncertainty in damping ratio. As for the priors, we choose uniform priors for the unknown parameters over ranges similar to those in \cite{vd2021}, additionally considering the statistical and seismological  results by \cite{aschwanden17,goddard17,pascoe18}:  $\mathcal{U}(R[{\rm Mm}], 0.5, 7)$, $\mathcal{U}(\zeta, 1.1, 6)$, and $\mathcal{U}(l/R, 0.1, 2)$.

Figure~\ref{fig:f1} shows the resulting marginal likelihoods for the two damping models over the grid of synthetic data in damping ratio and oscillation amplitude. The magnitude of the marginal likelihood at each point over the surface is a measure of how well a particular combination of damping ratio and oscillation amplitude is predicted by each model. This magnitude depends on the functional form of the damping models on their parameters and is not necessarily connected with their linear/nonlinear nature. Certain damping ratio and oscillation amplitude combinations are predicted more often than others. The predictions from each model clearly differ.

For nonlinear damping (Fig.~{\ref{fig:f1}, left panel),  the significant magnitudes of marginal likelihood are distributed over the triangular region with right angle in the lower left corner of the domain. The filled contour plot shows a convex structure with a triangular shape. The area with the highest marginal likelihood  corresponds to strong damping regimes with oscillation amplitudes in the range $\sim$ [5, 20] Mm. The obtained marginal likelihood distribution offers a straightforward explanation of the spreading of samples of pairs of oscillation amplitudes and damping ratios in the Monte Carlo analysis by \cite{vd2021}. In addition, it quantifies their relative plausibility.  For linear resonant absorption (Fig.~\ref{fig:f1}, right panel), the region with the highest marginal likelihood corresponds to low damping ratio values, independently of the oscillation amplitude. The predictive accuracy of the model decreases for weaker damping. 

\begin{figure*}[!t]
               \centering
         \includegraphics[scale=0.6]{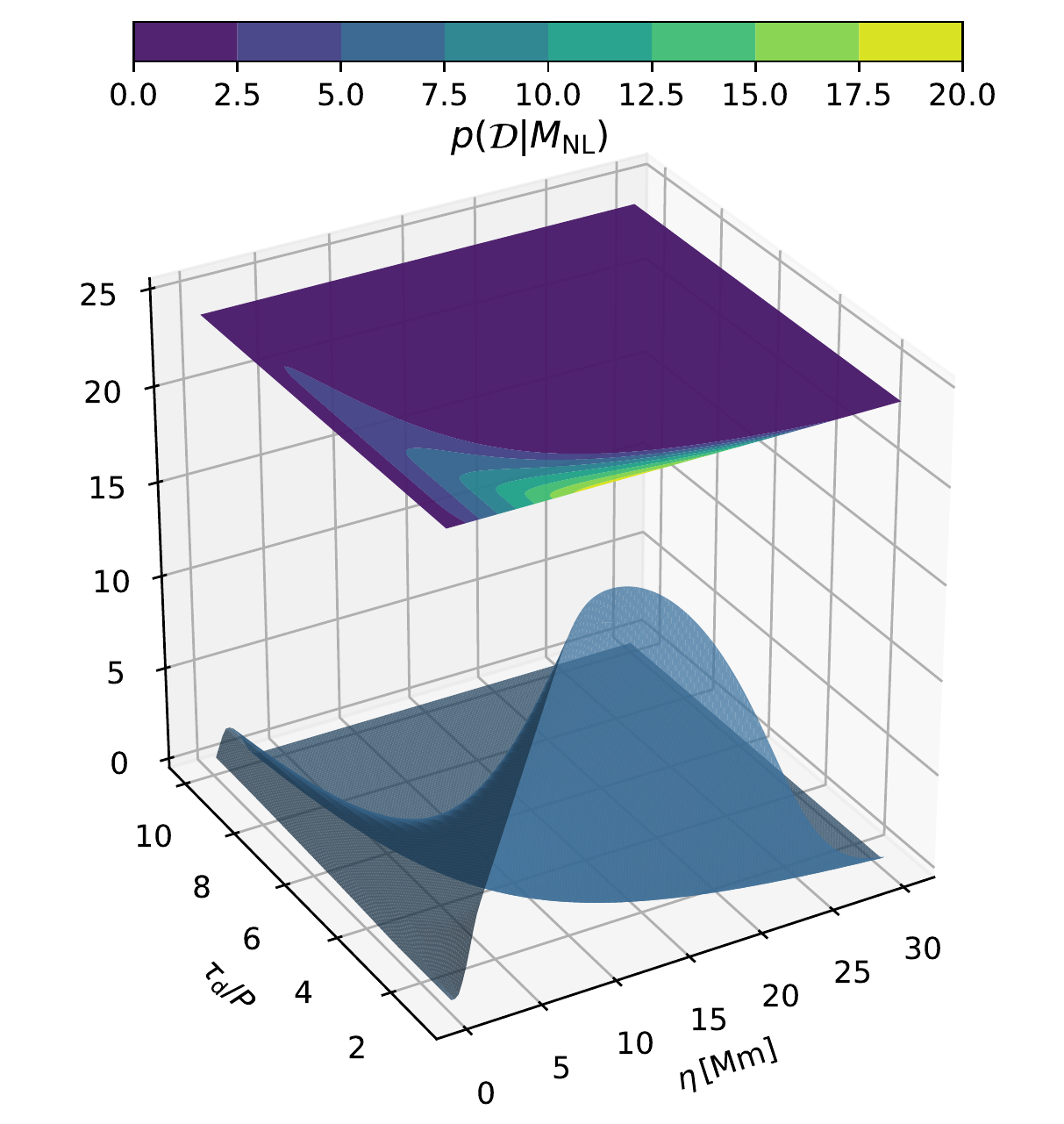}
          \includegraphics[scale=0.6]{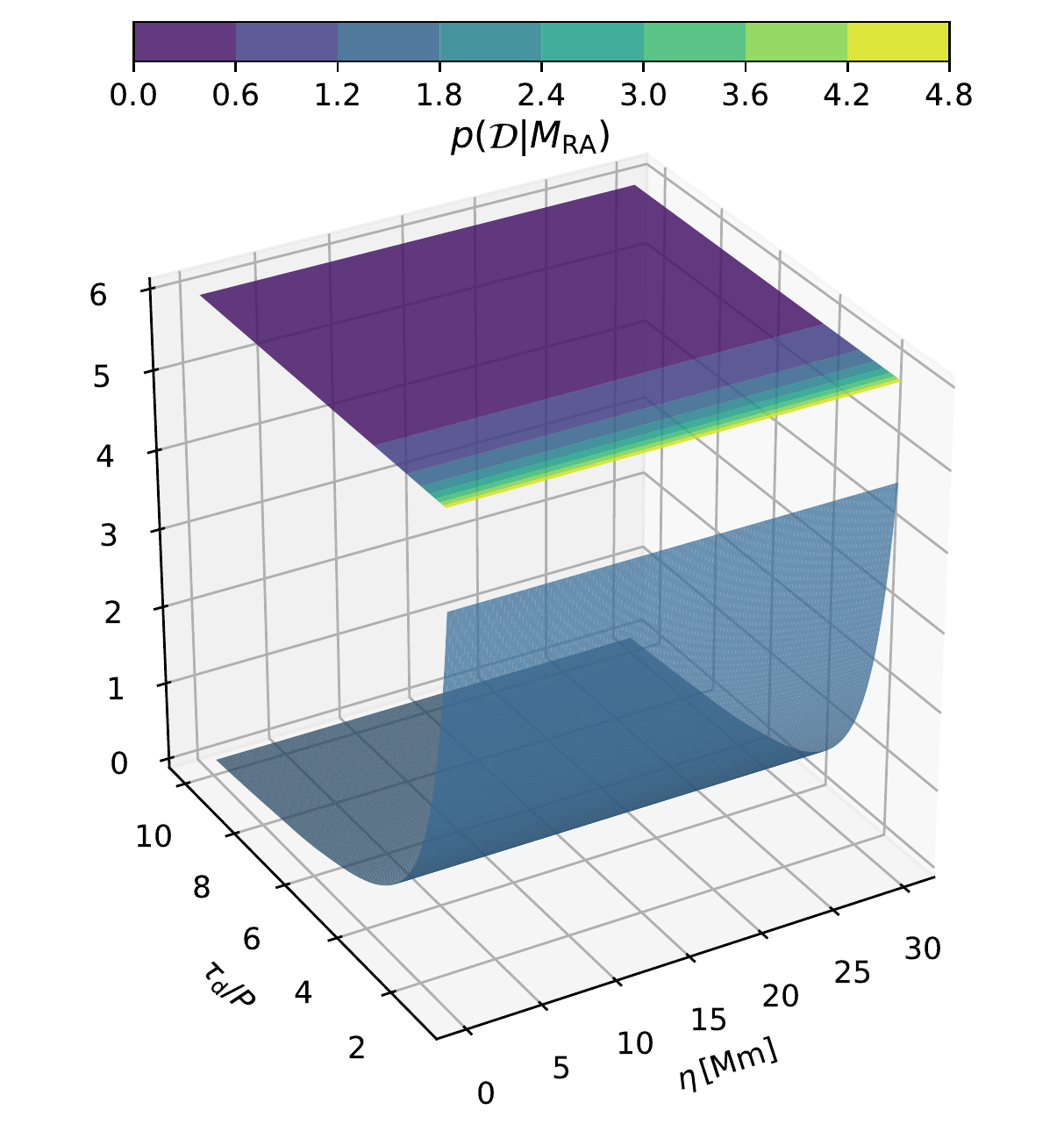}
\caption{Surface and filled contour representations of the marginal likelihood for the damping models $M_{\rm NL}$ (left panel) and $M_{\rm RA}$ (right panel) in the synthetic data space  $\mathcal{D}=(\eta, \tau_{\rm d}/P)$. Equation~(\ref{eq:ml}) is computed over a grid with $N_{\eta}=121$ and $N_{\tau_{\rm d}/P}=181$ points over the ranges $\eta\in[0.2-30]$ and $\tau_{\rm D}/P\in[1,10]$.   The error model in the likelihood function (\ref{eq:like}) is $\sigma= 0.1 \tau_{\rm d}/P$. \label{fig:f1}}
\end{figure*}

The marginal likelihood values are influenced by the quality of the data, the accuracy of the models and the prior assumptions.  The damping times predicted by the nonlinear model $M_{\rm NL}$ show certain differences with the numerical results of \cite{magyar16}. The resonant damping model $M_{\rm RA}$ describes the long term asymptotic state of long wavelength oscillations in tubes with thin boundaries. It overestimates the damping for thick layers \citep{vandoorsselaere04a,soler14a}. Linear MHD simulations by \cite{pascoe19} show that $M_{\rm RA}$ significantly overestimates the damping that would occur in the first few cycles of an oscillation when $l/R$ is not small, resulting in a model with an over preference for low damping ratio values. This means that $M_{\rm RA}$ has the largest marginal likelihood precisely in the strong damping regime in which it provides the least accurate description of the oscillations, which inevitably overemphasises the weakness of $M_{\rm RA}$ and hence the strength of $M_{\rm NL}$. Lack of knowledge about the radial density profile \citep{pascoe17b,arregui19} is an additional source of inaccuracy. Our prior sensitivity analysis showed that further increasing the upper limit for density contrast to $\zeta=9.5$ increases the magnitude of $p(\mathcal{D}|N_{\rm NL})$ and the marginal likelihood surface in Fig.~\ref{fig:f1} (left panel) extends towards combinations with slightly larger amplitude and low damping ratio values. Because larger loop radii correspond to a lower nonlinearity parameter $\eta/R$, increasing the upper limit for $R$ enables a greater range of $\tau_{\rm d}/P$ to be consistent with a particular observed amplitude.

Overall, there is qualitative agreement between the regions with high marginal likelihood for the nonlinear damping model and the broad location of observed data (compare Fig.~\ref{fig:f1} (left panel) with Figure 6 in \citealt{nechaeva19}). This is suggestive of relational evidence between the observations and the nonlinear damping model. 

The marginal likelihood only quantifies the evidence for a model in relation to the data that it predicts. In the general model comparison problem the aim is to assess the relative evidence between alternative models in explaining the same observed data. This is achieved with the calculation of the posterior ratio $p(M_{\rm NL}|\mathcal{D})/p(M_{\rm RA}|\mathcal{D})$. If the two models are equally probable a priori, $p(M_{\rm NL}) = p(M_{\rm RA})$ and by application of Bayes rule the posterior ratio reduces to the ratio of marginal likelihoods of the two models 

\begin{equation}\label{eq:bf}
B_{\rm NLRA} = 2\log \frac{p(\mathcal{D}|M_{\rm NL})}{p(\mathcal{D}|M_{\rm RA})} = -B_{\rm RANL},
\end{equation}
where the logarithmic scale is used for convenience in the evidence assessment.

The Bayes factors $B_{\rm NLRA}$ and $B_{\rm RANL}$ defined in Eq.~(\ref{eq:bf}) are a measure of relative evidence.  They quantify the relative plausibility of each of the two models to explain the same data. To assess the levels of evidence the empirical table by \cite{kass95} is employed. For instance, the evidence in favor of model $M_{\rm NL}$ in front of model $M_{\rm RA}$ is inconclusive for values of $B_{\rm NLRA}$ from 0 to 2; positive for values from 2 to 6; strong for values from 6 to 10; and very strong for values above 10. A similar tabulation applies to $B_{\rm RANL}$. 
 
Figure~\ref{fig:f2} shows the resulting Bayes factor distributions over the $\mathcal{D}$-space. By construction, regions in this space where $B_{\rm NLRA}$ and $B_{\rm RANL}$ reach the different levels of evidence are mutually exclusive and cannot overlap. There is a clear separation between the regions where $p(\mathcal{D}|M_{\rm NL}) > p(\mathcal{D}|M_{\rm RA})$, thus $B_{\rm NLRA}>0$, from those where  $p(\mathcal{D}|M_{\rm RA}) > p(\mathcal{D}|M_{\rm NL})$, hence $B_{\rm RANL}>0$. 

\begin{figure*}[!t]
               \centering
         \includegraphics[scale=0.6]{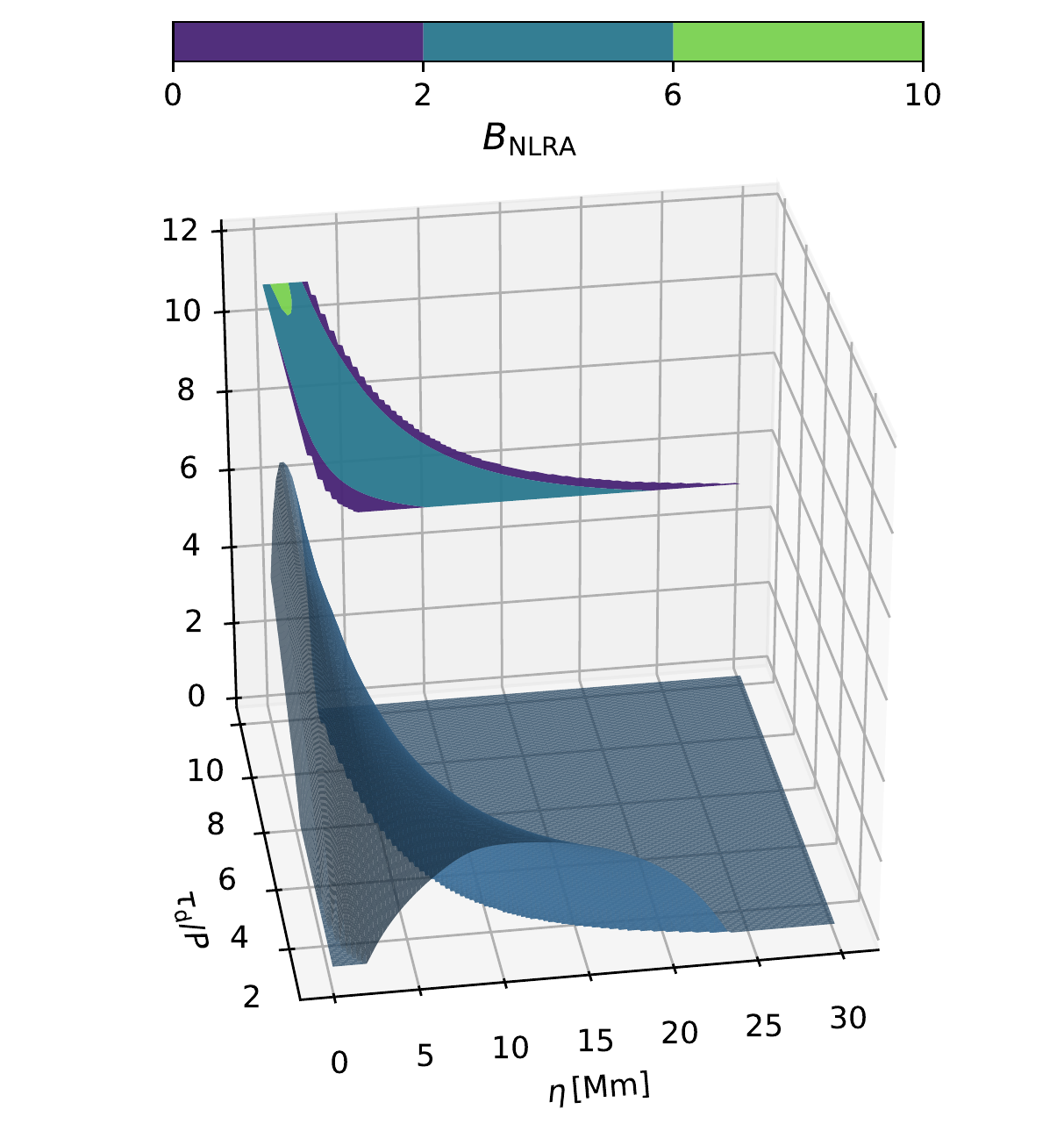}
         \includegraphics[scale=0.6]{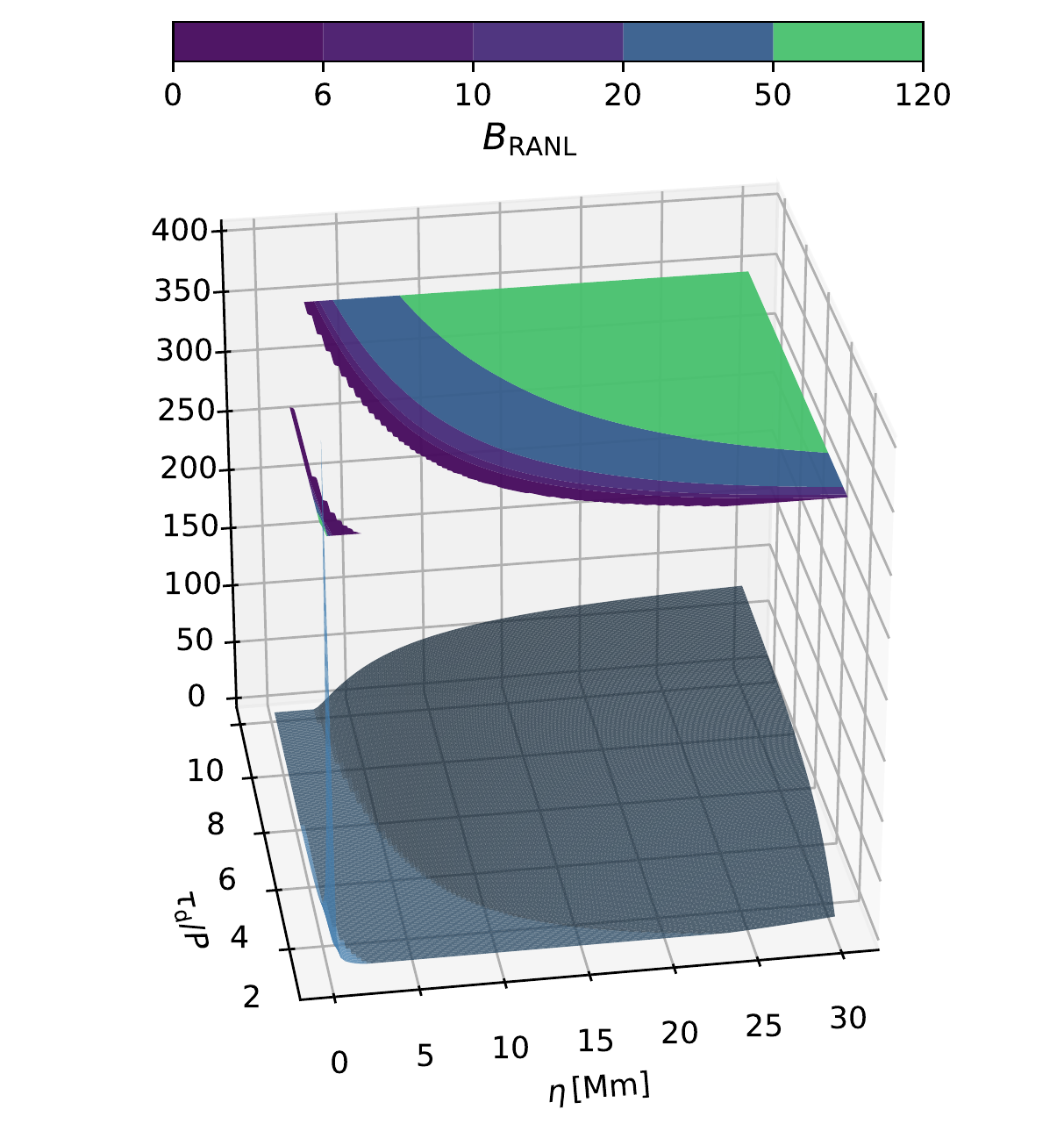}
\caption{Surface and filled contour representations of the Bayes factors $B_{\rm NLRA}$ (left panel) and $B_{\rm RANL}$ (right panel) in the synthetic data space $\mathcal{D}=(\eta, \tau_{\rm d}/P)$. Equation~(\ref{eq:bf}) is computed using the marginal likelihood calculations in Figure~\ref{fig:f1}.\label{fig:f2}}
\end{figure*}

Figure~\ref{fig:f2} (left panel) shows that the evidence supports nonlinear damping in a particular region of data space. The region contains combinations with small oscillation amplitudes in a narrow band below $\sim 5$ Mm and large damping ratio and extends towards combinations with smaller damping ratio and larger oscillation amplitudes in the broader range $\sim$ [2, 23] Mm. Because $p(\mathcal{D}|M_{\rm RA})$ is independent of oscillation amplitude, there is a conformity between the regions with high Bayes factor $B_{\rm NLRA}$ and marginal likelihood $p(\mathcal{D}|M_{\rm NL})$,  although their structures  differ. Figure~\ref{fig:f2} (right panel) shows that the evidence supports resonant absorption in two regions. The largest one extends towards the right-hand side of the domain. Here, the large Bayes factor values in favor of resonant damping are due to the corresponding low values of $p(\mathcal{D}|M_{\rm NL})$ relative to  $p(\mathcal{D}|M_{\rm RA})$ (see Fig.~\ref{fig:f1}). Because the Bayes factor represents relative strengths, one model being poor at a particular region in data space makes the alternative model seem better in comparison. The evidence in favor of resonant damping is very strong in the region restricted to combinations of very small amplitude oscillations with strong damping, close to the lower-left corner in data space.

Overall, there is qualitative agreement between the regions with high Bayes factor for the nonlinear damping model and the broad location of observed data (compare Fig.~\ref{fig:f2} (left panel) with Figure 6 in \citealt{nechaeva19}). This is suggestive of  evidence in favor of the nonlinear damping model relative to linear resonant absorption.

\subsection{Application to observations}\label{sec:app}

The catalog by \cite{nechaeva19} contains information about 223 oscillating loops observed by SDO/AIA at the 171 \AA\ extreme ultraviolet channel in the period 2010-2018 (see their Table 1). It is an extension of the catalog by \cite{goddard16a}. A subset of 101 cases contains information about both the damping ratio and the oscillation amplitude.  The methods described above to compute the marginal likelihood and the Bayes factor are applied to them to assess the strength of the evidence for the nonlinear damping model relative to that for resonant absorption. 

\begin{figure*}[!t]
         \centering
         \includegraphics[scale=0.6]{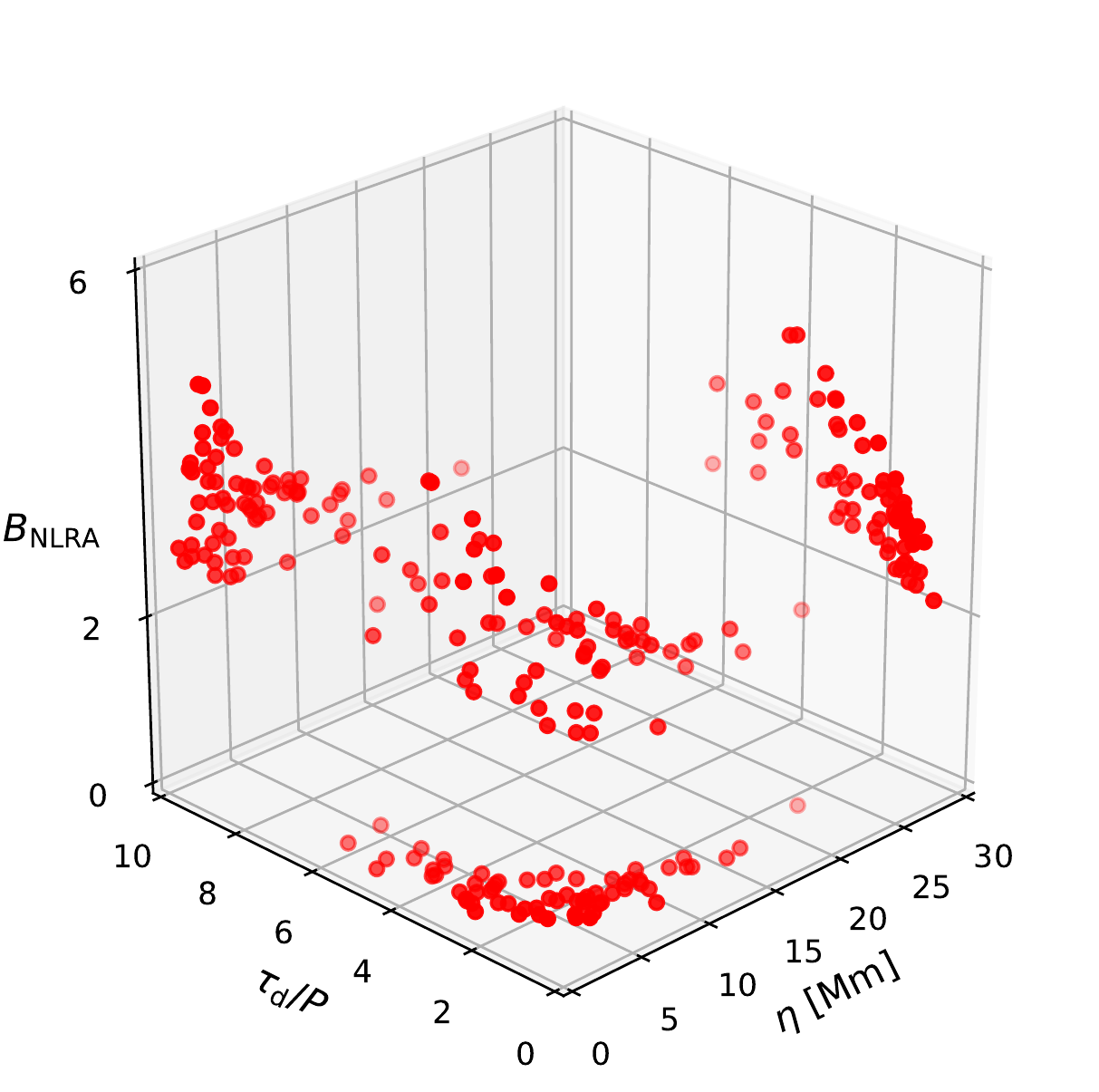}
         \includegraphics[scale=0.6]{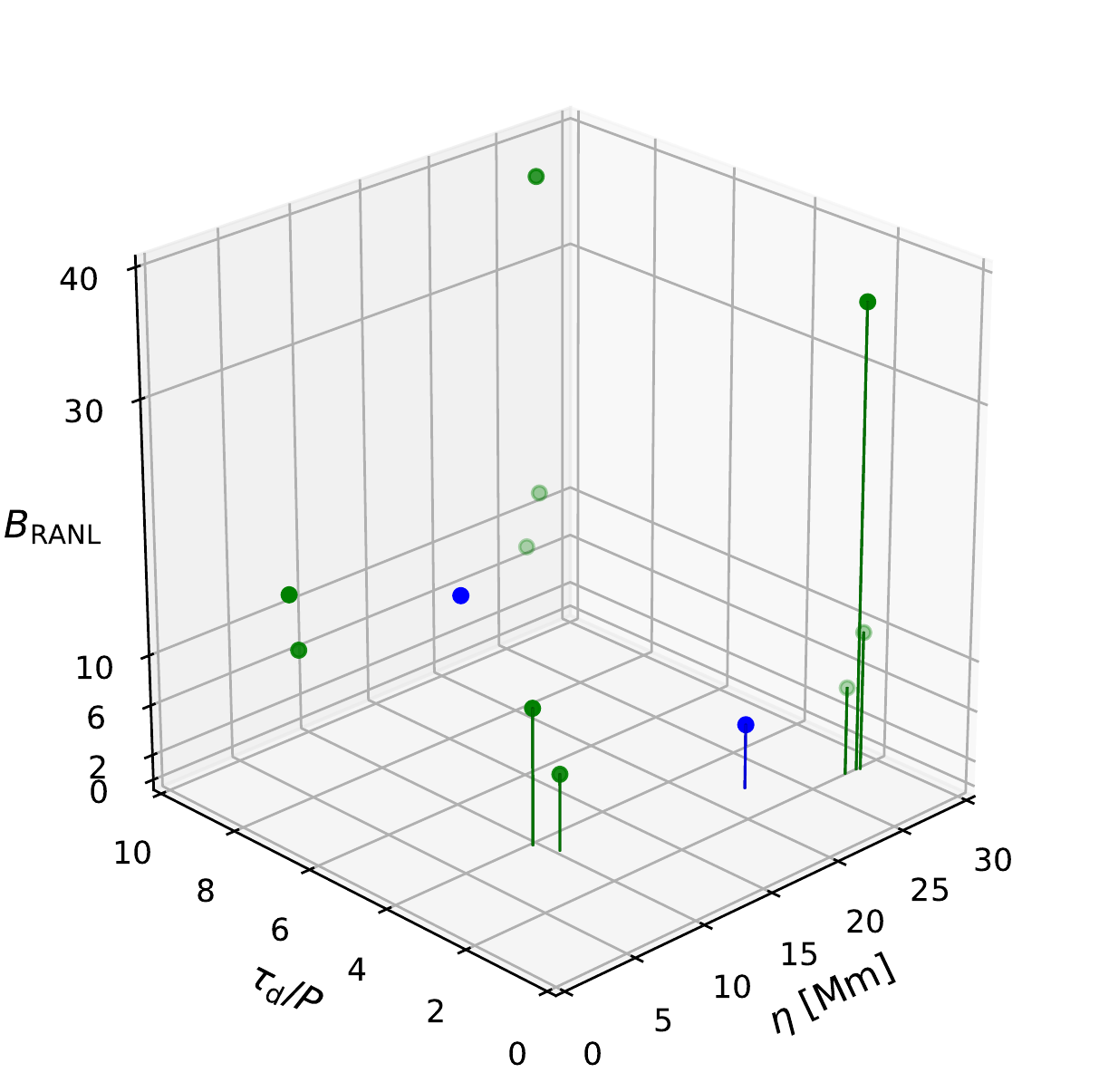}
\caption{Bayes factor, oscillation amplitude and damping ratio values for loop oscillations in Table~\ref{tab:nechaevadata} with positive evidence for nonlinear damping (left) and with positive (blue)/strong (green) evidence for resonant damping (right). \label{fig:f3}}
\end{figure*}

Table~\ref{tab:nechaevadata} in the Appendix shows a listing of the cases, with the event and loop identification numbers in \cite{nechaeva19}; the oscillation period; the damping time; the damping ratio; and the oscillation amplitude.  Instead of a grid over synthetic data space, the analysis now considers the specific pairs of measured values for oscillation amplitude and damping ratio for each event, $D=D_i =\{\eta_i, (\tau_{\rm d}/P)_{i}\}_{i =1}^{101}$. Equations~ (\ref{eq:ra}) and (\ref{eq:nl}) with the same prior assumptions as before are used to compute the predictions from the models. Predictions are confronted to observations in the computation of the likelihood function in Eq.~(\ref{eq:like}) and marginal likelihoods of the models using Eq.~({\ref{eq:ml}}). The ratio of marginal likelihoods in Eq.~(\ref{eq:bf}) gives the Bayes factor for each case. The results obtained for $B_{\rm NLRA}$ are given in the rightmost column of Table~\ref{tab:nechaevadata}. A positive value implies evidence in favor of the nonlinear damping model. A negative value implies evidence in favor of resonant absorption. The strength of the evidence is assessed with the empirical table by \cite{kass95}.

The marginal likelihood for the nonlinear damping model is larger than the marginal likelihood for resonant damping for the majority of cases,  91 out of 101 with $B_{\rm NLRA} > 0$. The opposite happens for the remaining 10 cases with $B_{\rm NLRA} < 0$ ( $B_{\rm RANL} > 0$). The level of evidence is conclusive in  71 cases, with Bayes factors above 2. The level of evidence is inconclusive in 30 cases,  with Bayes factors below 2. The number of cases with positive evidence for the nonlinear damping model is 65 ($2 < B_{\rm NLRA} < 6$).  There is a single case with positive evidence for resonant damping ($2 < B_{\rm RANL} < 6$). The number of cases with strong or very strong evidence for resonant damping is  5 ($B_{\rm RANL} > 6$).  The numerical values of $B_{\rm NLRA}$ depend on the chosen priors. Increasing the upper limit for $\zeta$ slightly favors $M_{\rm NL}$ while decreasing the upper limit for $R$ tends to favor $M_{\rm RA}$.

\begin{figure}[!t]
          \plotone{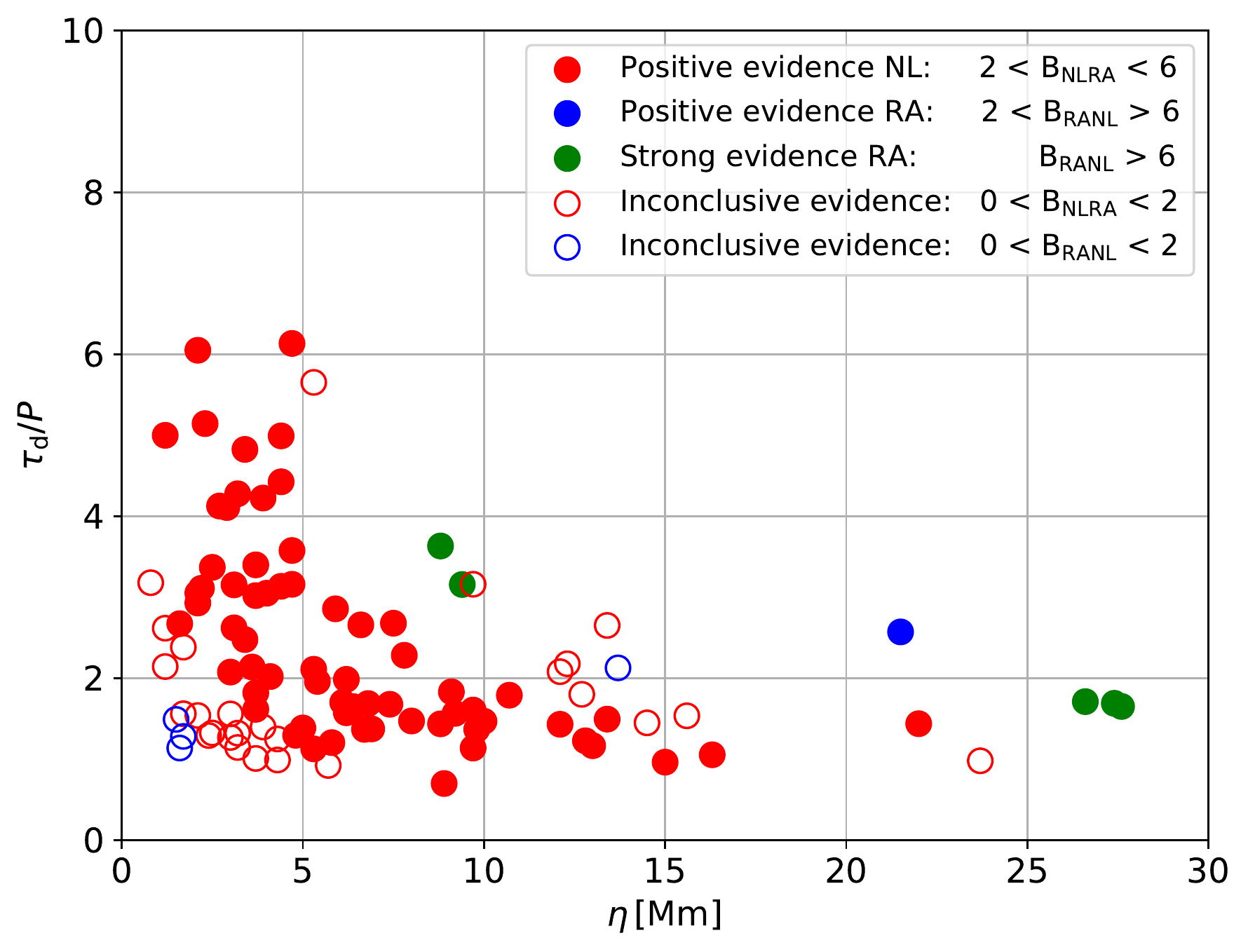}
          \caption{Scatter plot of oscillation amplitude and damping ratio values for the 101 loop oscillation cases in Table~\ref{tab:nechaevadata}. Color filled circles represent cases with conclusive evidence. Edge colored circles represent cases with either $p(\mathcal{D}|M_{\rm NL}) > p(\mathcal{D}|M_{\rm RA})$ or vice versa, yet inconclusive evidence.  Red is for $M_{\rm NL}$, blue and green for $M_{\rm RA}$. \label{fig:f4}}
\end{figure}

Figure~\ref{fig:f3} shows Bayes factor, oscillation amplitude, and damping ratio values for the cases with conclusive evidence supporting either the nonlinear damping model (left panel) or resonant absorption (right panel). The 65 cases with positive evidence for the nonlinear model are grouped in the region in data space where the previous analysis of the structure of the evidence favored nonlinear damping. The 6 cases with positive or strong evidence for resonant damping are distributed in the complementary region where the low marginal likelihood for nonlinear damping favors resonant absorption.  

Figure~\ref{fig:f4} shows the results obtained for all 101 loop oscillations, regardless of the conclusive or inconclusive nature of the evidence. The symbols and their colors indicate the corresponding levels of evidence. The pattern of evidence distribution in the general structure analysis in Sect.~\ref{sec:structure} is maintained. Among the cases with inconclusive evidence, those with larger marginal likelihood for nonlinear damping (edge colored red circles) are located at the two sides of the main area with positive evidence for nonlinear damping. Those with larger marginal likelihood for resonant damping (edge colored blue circles), spread further to the sides,  with three of them corresponding to combinations of  small amplitudes with strong damping.

The amplitudes in the observational data may be consistently underestimated, depending on the orientation of the loop and the distance of the observational slit from the loop apex, because they represent the apparent amplitude projected in the plane of the sky. Since 
the range of damping ratio values which can be accurately described by $M_{\rm NL}$ decreases for larger amplitudes (Fig.~\ref{fig:f1}, left panel),  the effect of underestimating the actual amplitude would generally benefit this model, while having no effect on $M_{\rm RA}$.

\section{Conclusion and discussion}

Our judgement about the degree of belief in a particular model to explain observed data is based on the assessment of how well the model predictions agree with observed data and on how much better this is achieved in comparison to alternative models. These relational and relative measures of the evidence for a model are given by the marginal likelihood and the Bayes factor, respectively. We considered two models postulated to explain the damping of transverse oscillations in solar coronal loops and applied those principles of Bayesian reasoning. 

The structure of the evidence shows a clear separation between the regions in data space where the nonlinear damping model is more plausible relative to the linear resonant absorption model and vice versa. There is qualitative agreement between the regions of high marginal likelihood and Bayes factor for the nonlinear damping model and the location of measured data for damping ratio and oscillation amplitude. The evidence is quantified by application to 101 loop oscillations observed with SDO/AIA. The marginal likelihood for the nonlinear damping model is larger in the majority of the cases and conclusive evidence in favor of this model is obtained in 65 cases. The cases with conclusive evidence for the nonlinear damping model outnumber those in favour of linear resonant damping  by a factor of 10. The evidence for the nonlinear damping model relative to linear resonant absorption is therefore appreciable to a reasonable degree of Bayesian certainty.

The Bayes factor values in this study cannot be regarded in absolute terms. Their accuracy is determined by the quality of the data, the level of refinement of the models, and the more or less informative nature of the prior assumptions. Improvement on the evidence assessment can be achieved, for example, with the use of alternative prescriptions for linear resonant absorption that more accurately describe the early stage of loop oscillations, as appropriate for the observational cases of strong damping \citep{pascoe19,hood13}, the consideration of additional physical effects \citep{pascoe17b}, or by further constraining the prior densities through seismological inference of the unknown parameters \citep{nistico13,pascoe16,pascoe17a}.

The availability of specific model parameters also affects the calculated Bayes factors. As an example, consider entry number i=26 in Table~\ref{tab:nechaevadata}, with $B_{\rm NLRA}=-10.9$. For this event, first studied by  \cite{nistico13}, estimates for loop radius, $R\sim3.3$ Mm; density contrast, $\zeta\sim3$; and thickness of the layer, $l/R\sim0.5$, are available \citep{pascoe16,pascoe17a,pascoe17b}. In contrast to the value of $\eta=8.8$ Mm for the amplitude reported in the catalog, \cite{pascoe17b} quote a value of $\eta\sim 2$ Mm. Bayes factor calculations, now without the need to perform integration over parameter space, lead to values of $B_{\rm NLRA}\sim13$ for $\eta = 2.2$ Mm and $B_{\rm NLRA}\sim-75$ for $\eta = 8.8$ Mm, which give support to the nonlinear model and also point to an overestimation of the amplitude in the catalog.

In general,  because of our inability to directly measure the relevant physical parameters, the Bayesian approach offers the best course of action for judgment under uncertainty and offers principled ways to improve the evidence assessment as better data, more refined models and more informative priors become available.


\acknowledgments
This research was supported by project PGC2018-102108-B-I00 from Ministerio de Ciencia, Innovaci\'on y Universidades and FEDER funds. 

\facilities{SDO/AIA \citep{lemen12}}
\software{NumPy \citep{numpy20},  Matplotlib \citep{hunter07}}


\begin{appendix}\label{sec:append}
Table with the results of the analysis described in Section~\ref{sec:app}.

\startlongtable
\begin{deluxetable*}{cccccccc}
\tablenum{1}
\tablecaption{Loop oscillation data from \cite{nechaeva19} and computed Bayes factors. \label{tab:nechaevadata}}
\tablewidth{0pt}
\tablehead{
\colhead{$i$} & \colhead{Event ID} & \colhead{Loop ID} & \colhead{P [min]}  &
 \colhead{$\tau_{\rm D}$ [min]}  & \colhead{$\tau_{\rm D}/P$} & \colhead{$\eta$ [Mm]} & \colhead{$B_{\rm NLRA}$}
}
\startdata
1 & 1  &  1  &  3.42 $\pm$ 0.06  &      5.34 $\pm$ 1.12 &1.56 $\pm$ 0.33& 1.7 &0.7\\
2 & 1  &  2  &  4.11 $\pm$ 0.05  &    10.76 $\pm$ 2.79 &2.62 $\pm$ 0.68& 1.2 &1.5 \\
3 & 3  &  1  &  2.46 $\pm $ 0.03  &      8.8 $\pm$ 1.8 &3.58 $\pm$ 0.73& 4.7&3.9\\  
4 & 3  &  2  &  3.62 $\pm$ 0.08  &      4.12 $\pm$  0.47 &1.14 $\pm$ 0.13& 9.7&2.8\\ 
5 & 4  &  1  &  2.29 $\pm$ 0.03  &       7.18 $\pm$  1.5 &3.14 $\pm$ 0.66& 4.4&3.9\\ 
6 & 4  &  2 &   3.47 $\pm$ 0.03  &         7.44 $\pm$ 1.& 2.14 $\pm$ 0.29& 1.2&1.2\\   
7 & 7  &  1 &   1.69 $\pm$ 0.02  &        7.23 $\pm$  1.3 & 4.28 $\pm$ 0.77& 3.2& 4.5\\  
8 & 8  &  1 &   3.74 $\pm$ 0.07  &        10.  $\pm$  1.& 2.67 $\pm$ 0.27&  1.6&2.5\\   
9 & 9  &  1 &   5.14 $\pm$ 0.17  &        5.09 $\pm$ 0.98&0.99 $\pm$ 0.19 & 4.3& 1.3\\ 
10 & 9  &  2 &   8.95 $\pm$ 0.14  &        11.83  $\pm$ 4.76& 1.32 $\pm$ 0.53& 3.2& 1.4\\ 
11 & 10 &  1 & 11.46 $\pm$ 0.17  &        8.02 $\pm$  1.09&0.70 $\pm$ 0.10 &  8.9&2.0\\ 
12 & 11 &  3 &    2.6 $\pm$ 0.05   &         8.84 $\pm$ 1.5& 3.40 $\pm$ 0.58&  3.7&4.2\\  
13 & 14 &  2 &  2.35 $\pm$ 0.07  &         2.69 $\pm$ 0.64& 1.14 $\pm$ 0.27&  3.2&1.0\\ 
14 & 15 &  1 &  2.07 $\pm$ 0.04  &        9.99 $\pm$ 4.59&4.83 $\pm$ 2.22 &   3.4&3.4\\
15 & 16 &  2 &  9.52 $\pm$ 0.11  &         12.2 $\pm$  3.47& 1.28 $\pm$ 0.36&  1.7&-0.1\\ 
16 & 17 &  2 & 11.27 $\pm$ 0.12 &       16.55 $\pm$  1.44& 1.47 $\pm$ 0.13&  10.0&3.0\\ 
17 & 18 &  1 &  5.36 $\pm$ 0.23  &        16.19 $\pm$ 7.67& 3.02 $\pm$ 1.44&  3.7&3.1\\ 
18 & 21 &  1 & 15.36 $\pm$ 0.4  &        19.19 $\pm$ 1.55&1.25 $\pm$ 0.11 &   4.3&2.0\\
19 & 22 &  1 & 17.86 $\pm$ 0.3  &      27.43 $\pm$ 4.26&1.54 $\pm$ 0.24 &  15.6&0.7\\ 
20 & 22 &  3 & 20.46 $\pm$ 0.58 &       35.01 $\pm$  6.44&1.71 $\pm$ 0.32 &   26.6&-7.0\\
21 & 23 &  1 &  5.13 $\pm$  0.11 &         8.  $\pm$    5.& 1.56 $\pm$ 0.98&   3.0&1.8\\   
22 & 24 &  1 & 11.95 $\pm$ 0.13  &       18.71 $\pm$ 4.5& 1.57 $\pm$ 0.38&   9.2&2.9\\ 
23 & 26 &  1 &  3.71 $\pm $ 0.05  &       7.83 $\pm$  0.62& 2.11 $\pm$ 0.17&   5.3&3.6\\
24 & 27 &  1 &  7.67 $\pm$  0.04  &       24.22 $\pm$ 2.02& 3.16 $\pm$ 0.26&   9.4&-6.2\\
25 & 27 &  2 &  9.59 $\pm$ 0.09   &       17.57 $\pm$  2.35& 1.83 $\pm$ 0.25&   9.1&3.0\\ 
26 & 28 &  1 &  4.28 $\pm$ 0.02   &       15.55 $\pm$ 1.22& 3.63 $\pm$ 0.29&  8.8& -10.9\\
27 & 28 &  2 &  3.38 $\pm$ 0.02   &       19.11 $\pm$ 4.85& 5.65 $\pm$ 1.44&  5.3& 1.7\\
28 & 30 &  1 &  9.95 $\pm$ 0.27   &       16.7 $\pm$  1.03& 1.68 $\pm$ 0.11&   7.4&3.3\\ 
29 & 34 &  2 &  5.2  $\pm$  0.08   &       15.23 $\pm$ 5.5& 2.93 $\pm$ 1.06&  2.1&2.7\\  
30 & 36 &  2 & 5.61 $\pm$  0.03   &        24.83 $\pm$  3.41& 4.43 $\pm$ 0.61& 4.4& 3.8\\ 
31 & 36 &  4 & 5.53 $\pm$  0.04   &        7.32 $\pm$ 1.08&1.32 $\pm$ 0.20 &  2.5&1.0\\ 
32 & 36 &  7 & 5.72 $\pm$  0.06   &       14.17 $\pm$  2.73&2.48 $\pm$ 0.48 &  3.4&3.5\\ 
33 & 36 &  8 & 4.33 $\pm$  0.08   &       9.01 $\pm$  2.16& 2.08 $\pm$ 0.50&   12.1&1.3\\
34 & 36  & 9 & 6.18 $\pm$  0.05   &      13.15 $\pm$ 2.66& 2.13 $\pm$ 0.43&  13.7&-0.4\\ 
35 & 37  & 1 & 7.14 $\pm$ 0.07    &       7.53 $\pm$ 1.45& 1.05 $\pm$ 0.20&   16.3& 2.3\\
36 & 37  & 2 & 3.6  $\pm$  0.03    &       9.44 $\pm$ 0.92& 2.62 $\pm$ 0.26&  3.1& 3.8\\
37 & 37  & 3 & 8.35 $\pm$ 0.08    &      15.04 $\pm$ 1.81& 1.80 $\pm$ 0.22&  12.7&1.2\\ 
38 & 37  & 5 & 4.5   $\pm$ 0.02    &       14. $\pm$   2.& 3.11 $\pm$ 0.44&   2.2&3.7\\   
39 & 38  & 1 & 7.23 $\pm$ 0.06    &       15.75 $\pm$  3.09& 2.18 $\pm$ 0.43&  12.3&0.4\\ 
40 & 38  & 2 & 9.78 $\pm$ 0.19    &       14.62 $\pm$ 4.96& 1.49 $\pm$ 0.51&   13.4&2.3\\ 
41 & 38  & 3 & 6.95 $\pm$ 0.14    &         9. $\pm$   3. & 1.29 $\pm$ 0.43&  4.8&2.1\\   
42 & 39  & 1 & 2.48 $\pm $0.04    &         7.82 $\pm$ 1.66& 3.15 $\pm$ 0.67&  3.1&4.0\\ 
43 & 42  & 1 & 15.28 $\pm$ 0.16  &       21.98 $\pm$ 15.6& 1.44 $\pm$ 1.02&  22.0&2.3\\  
44 & 42  & 2 & 15.76 $\pm$  0.12 &       26.64 $\pm$   2.17& 1.69 $\pm$ 0.14&   27.4&-36.8\\
45 & 42  & 3 & 16.08 $\pm$  0.21 &       15.76 $\pm$   3.09& 0.98 $\pm$ 0.19&   23.7&0.8\\
46 & 43  & 4 & 10.45 $\pm$  0.17 &       15.38 $\pm$   2.58& 1.47 $\pm$ 0.25&   8.0&3.1\\
47 & 43  & 5 & 8.03 $\pm$  0.18   &      9.37 $\pm$   1.22& 1.17 $\pm$ 0.15&   13.0&2.7\\
48 & 46  & 5 & 4.8  $\pm$  0.1      &       19.72 $\pm$   3.23& 4.11 $\pm$ 0.68&   2.9&4.5\\
49 & 48  & 1 & 9.07 $\pm$  0.14   &       20.71 $\pm$   4.71& 2.28 $\pm$ 0.52&   7.8&3.0\\
50 & 48  & 2 & 11.88  $\pm$ 0.13 &      19.62 $\pm$   2.96& 1.65 $\pm$ 0.25&   27.6&-11.1\\
51 & 48  & 5 & 13.5  $\pm$  0.16  &      24.17 $\pm$   5.13& 1.79 $\pm$ 0.38&   10.7&2.5\\
52 & 48  & 7 & 14.16 $\pm$  0.55 &      13.64 $\pm$   3.93& 0.96 $\pm$ 0.28&   15.0&2.7\\
53 & 49  &  1 & 8.52 $\pm$  1.02  &  8.6 $\pm$  3.6 & 1.01 $\pm$ 0.44&  3.7    $\pm$ 1.7 &1.2\\  
54 & 50  &  1 & 7.36 $\pm$  0.26  &   24.8 $\pm$ 12.1 & 3.37 $\pm$ 1.65&  2.5   $\pm$  0.6 &3.0\\  
55 & 51  &  2 & 13.75 $\pm$  0.38 &   23.2 $\pm$   8.9 & 1.69 $\pm$ 0.65&   6.8   $\pm$ 1.5  &2.9\\ 
56 & 52  &  1 & 13.51 $\pm$  0.42 &   27.3 $\pm$  11.1 & 2.02 $\pm$ 0.82&  4.1  $\pm$ 0.8   &2.7\\ 
57 & 52  &  2 & 12.95  $\pm$ 0.73  &    15.6 $\pm$  4.4 & 1.20 $\pm$ 0.35&  5.8 $\pm$ 1.5   &2.3\\ 
58 & 53  &  1 & 16.17 $\pm$  0.42  &    43.0 $\pm$  13.5 & 2.66 $\pm$ 0.84&  6.6 $\pm$ 1.3   &3.1\\ 
59 & 54  &  1 & 11.16 $\pm$  0.86  &    18.4 $\pm$   8.2 & 1.65 $\pm$ 0.75&  6.4  $\pm$ 1.6   &2.8\\
60 & 56  &  3 & 11.83 $\pm$  0.38  &    31.7 $\pm$  10.6 & 2.68 $\pm$ 0.90& 7.5  $\pm$ 1.3    &2.7\\
61 & 57  &  2 & 2.89 $\pm$  0.09  &      6.0 $\pm$   1.8 & 2.08 $\pm$ 0.63& 3.0   $\pm$ 0.8    &2.5\\
62 & 58  &  1 & 3.44 $\pm$  0.46  &      8.2 $\pm$   4.6 & 2.38 $\pm$ 1.37&  1.7   $\pm$ 1.0   &1.5\\
63 & 59  &  1 & 5.19 $\pm$  0.33  &    11.1 $\pm$   5.2 & 2.14 $\pm$ 1.01& 3.6   $\pm$  1.2    &2.6\\
64 & 60  &  3 & 9.59 $\pm$  0.66  &    27.4 $\pm$  12.5 & 2.86 $\pm$ 1.32&  5.9   $\pm$  2.2   &2.9\\
65 & 63  &  1 & 4.45 $\pm$  0.07  &    27.3 $\pm$  10.2 & 6.13 $\pm$ 2.29& 4.7   $\pm$  0.9   &2.6\\
66 & 64  &  1 & 2.17 $\pm$  0.13  &     6.9 $\pm$   3.6 & 3.18 $\pm$ 1.67& 0.8   $\pm$  0.3    &0.5\\
67 & 64  &  3 & 7.19 $\pm$  0.20  &    35.9 $\pm$  14.1 & 4.99 $\pm$ 1.97&  4.4   $\pm$  0.5   &3.1\\
68 & 64  &  5 & 20.76 $\pm$  1.03  &     28.4 $\pm$   7.3 & 1.37 $\pm$ 0.36&   9.8$\pm$  2.4  &2.8\\
69 & 64  &  6 & 20.43 $\pm$  0.47  &    29.6 $\pm$   7.5 & 1.45 $\pm$ 0.37&  14.5$\pm$ 2.6   &1.9\\
70 & 65  &  2 & 17.82 $\pm$  0.49  &     25.6 $\pm$   6.9 & 1.44 $\pm$ 0.39&  8.8 $\pm$ 2.0   &2.9\\
71 & 66  &  1 & 2.38 $\pm$   0.05  &     14.4 $\pm$   7.0 & 6.05 $\pm$ 2.94&  2.1  $\pm$ 0.4   &3.6\\
72 & 67  &  1 & 17.19 $\pm$  0.76 &    21.1 $\pm$   5.8 & 1.23 $\pm$ 0.34&   12.8 $\pm$ 3.6  &2.6\\
73 & 68  &  1 & 7.34 $\pm$   0.66  &    11.3 $\pm$   5.0 & 1.54 $\pm$ 0.70&  2.1  $\pm$  0.8    &0.9\\
74 & 69  &  1 & 6.22 $\pm$   0.18  &    32.0 $\pm$  14.8 & 5.14 $\pm$ 2.38&  2.3  $\pm$   0.5   &3.5\\
75 & 69  &  2 & 12.81 $\pm$  0.32  &   21.8 $\pm$   4.6 & 1.70 $\pm$ 0.36&   6.1  $\pm$ 1.3   &3.1\\
76 & 70  &  2 & 9.85 $\pm$  0.40  &     15.5 $\pm$   4.3 & 1.57 $\pm$ 0.44& 6.2  $\pm$   1.7    &2.9\\
77 & 70  &  3 & 5.56 $\pm$  0.16  &    27.8 $\pm$  13.7 & 5.00 $\pm$ 2.47&   1.2  $\pm$    0.3  &2.7\\
78 & 72  &  1 & 13.30 $\pm$  0.72  &   18.2 $\pm$   5.9 & 1.37 $\pm$ 0.45&  6.7  $\pm$ 1.8    &2.7\\
79 & 72  &  2 & 6.82 $\pm$   0.15  &    20.8 $\pm$   6.4 &    3.05 $\pm$ 0.94& 4.0  $\pm$  0.7 &3.6\\
80 & 72  &  4 & 18.59 $\pm$  0.78 &    49.3 $\pm$  20.6 & 2.65 $\pm$ 1.11&   13.4 $\pm$  2.1  &1.4\\
81 & 73  &  1 & 10.21 $\pm$  0.48  &   16.5 $\pm$   4.6 & 1.62 $\pm$ 0.46&  3.7  $\pm$ 0.9    &2.2\\
82 & 74  &  2 & 4.85 $\pm$  0.23  &    14.8 $\pm$   6.8 & 3.05 $\pm$ 1.41&   2.1  $\pm$   0.6   &2.6\\
83 & 74  &  3 & 2.90 $\pm$  0.27  &     3.3 $\pm$   1.0 & 1.14 $\pm$ 0.36&    1.6  $\pm$  0.7  &-0.7\\
84 & 75  &  1 & 10.76 $\pm$  0.28 &     15.4 $\pm$   3.8 & 1.43 $\pm$ 0.36&   12.1 $\pm$  1.1  &2.5\\
85 & 75  &  2 & 8.48 $\pm$  0.16  &    26.8 $\pm$   7.6 & 3.16 $\pm$ 0.90&    9.7   $\pm$  1.2  &1.0\\
86 & 75  &  3 & 8.71 $\pm$  0.42  &     17.3 $\pm$   5.4 & 1.99 $\pm$ 0.63&  6.2   $\pm$ 1.5   &3.1\\
87 & 77  &  1 & 7.69 $\pm$  0.50  &     24.3 $\pm$  12.8 & 3.16 $\pm$ 1.68&  4.7   $\pm$  1.4   &3.0\\
88 & 78  &  2 & 5.88 $\pm$  0.43  &     7.6 $\pm$   2.7 & 1.29 $\pm$ 0.47&   2.4   $\pm$  1.1   &0.7\\
89 & 79  &  3 & 2.95 $\pm$  0.33  &     4.4 $\pm$   2.2 & 1.49 $\pm$ 0.76&  1.5    $\pm$  0.8   &-0.0\\
90 & 80  &  1 & 12.51 $\pm$  0.57  &   20.1 $\pm$   5.5 & 1.61 $\pm$ 0.45&    9.7 $\pm$  2.5  &2.8\\
91 & 80  &  2 & 9.69 $\pm$  0.26  &     19.0 $\pm$   4.8 & 1.96 $\pm$ 0.50&   5.4  $\pm$   1.4  &3.2\\
92 & 81  &  1 & 11.35 $\pm$  0.83  &    14.4 $\pm$   5.1 & 1.27 $\pm$ 0.46&  3.0  $\pm$  1.1   &1.2\\
93 & 83  &  1 & 5.99 $\pm$   0.51  &    10.9 $\pm$   4.6 & 1.82 $\pm$ 0.78&  3.7  $\pm$   1.2   &2.4\\
94 & 84  &  1 &  3.78 $\pm$  0.33  &    15.6 $\pm$   8.3 & 4.13 $\pm$ 2.23&  2.7 $\pm$  1.0   &3.2\\
95 & 87  &  1 & 5.94 $\pm$   0.32  &     25.1 $\pm$  10.8 & 4.23 $\pm$ 1.83&   3.9  $\pm$ 1.0  &3.3\\
96 & 87  &  2 & 9.24 $\pm$   0.62  &   12.7 $\pm$   4.6 & 1.37 $\pm$ 0.51&    6.9  $\pm$  2.2  &2.7\\
97 & 88  &  1 & 9.38 $\pm$   0.19  &   13.0 $\pm$   5.4 & 1.39 $\pm$ 0.58&   5.0   $\pm$ 0.4   &2.4\\
98 & 88  &  3 & 13.75 $\pm$ 1.18  &     15.5 $\pm$   6.5 & 1.13 $\pm$ 0.48&  5.3  $\pm$  1.8   &2.1\\
99 & 90  &  1 &  9.32 $\pm$   0.31  &      8.6 $\pm$   2.4 & 0.92 $\pm$ 0.26&  5.7  $\pm$ 1.4   &1.8\\
100 & 92  &  1 & 6.51 $\pm$  0.31  &      9.1 $\pm$   2.4 & 1.40 $\pm$ 0.37&  3.9 $\pm$ 1.3    &2.0\\
101 & 93  &  1 & 8.32 $\pm$  0.10  &    21.4 $\pm$   4.9 & 2.57 $\pm$ 0.59&  21.5  $\pm$ 2.4   &-5.2\\
\enddata
\tablecomments{The table gives the entry number $i$, Event ID, Loop ID, oscillation period $P$, damping time $\tau_{\rm D}$, damping rate $\tau_{\rm D}/P$, oscillation amplitude $\eta$, and computed Bayes factor B$_{\rm NLRA}=-B_{\rm RANL}$. Errors for damping ratio are calculated from those for period and damping time as $R_{\tau_{\rm d}/P}=\sqrt{R^2_{\tau_{\rm d}} + R^2_{P}}$, where $R$ refers to the relative error.}
\end{deluxetable*}

\end{appendix}
\end{document}